\documentclass{article}
\usepackage[utf8]{inputenc}
\usepackage{geometry}
\bibliography{references}
\usepackage{graphicx}
\graphicspath{{images/}}

\title{Enhancing User Engagement in E-commerce through Dynamic Animations}
\author{Waaridh Borpujari \\ MIT ADT University \\ ADT23SOCB1333@students.mituniversity.edu.in}
\date{February 2025}

\begin{document}
\maketitle

\begin{abstract}

The use of animation to gain user attention has been increasing, supported by various studies on user behavior and psychology. However, excessive use of animation in interfaces can negatively impact the user. This paper deals with a specific type of animation within a specialized domain of e-commerce. Drawing upon theories such as the Zeigarnik Effect, Aesthetic-Usability effect, Peak-End rule, and Hick's law, we analyze user behavior and psychology when exposed to a dynamic price-drop animation. Unlike conventional static pricing strategy, this animation introduces movement to signify price reduction. In our theoretical study approach, we evaluate and present a user study on how such an animation influences user perception, psychology, and attention. If acquired effectively, dynamic animations can enhance engagement, spark anticipation, and subconsciously create a positive experience by reducing cognitive load.  

\end{abstract}

\section{Introduction}

Balancing value and aesthetics is a complex challenge in user experience design. The e-commerce industry continues to refine its psychological strategies to influence purchasing decisions. One such method is the display of discounted prices where the original price is shown alongside the reduced price in a static format. While effective, this method may not prove to fully engage users or maximize the psychological impact on users which is usually expected to take place. \newline Several studies show that well-designed UI animations enhance user experience, attention, and emotional engagement. However, excessive use of animation or poorly executed animations can overwhelm users particularly with cognitive sensitivities. This paper explores an alternative approach to presenting discounted prices: a dynamic price-drop animation where the price reduction is shown progressively rather than displayed statically. 

While the animation part does help in reducing cognitive load for the users, it is also necessary to consider that displaying a dynamic price drop animation for all the products can rather affect negatively than positively. If such animations are only displayed on the product page, it would pose either of the problems: 
\begin{enumerate}
\item Loss of Anticipation: if the discounted rate is already displayed on the Search Results page, the price drop animation will not create anticipation for the user.\footnote{if the sale price in the search results page is hidden, it would mean that the user would have to visit every single product page just to see its price which affects usability altogether.}
\item Feeling overwhelmed: if there are multiple things happening at a time on a single page, it confuses and overwhelms the user than doing any good.
\end{enumerate}

\newpage
Hence, this method should only be preferably employed for selected products which will be heavy on discounted price to create even more anticipation and respect the laws of making a simple interface for the user.

\subsection{User Set}

When designing apps for users, the awareness of types of users who will be using the app, should be considered as well because experiences differ from user to user. Thus, we analyse this approach through established psychological theories such as:

\begin{enumerate}

\item Aesthetic-usability Effect
\item Peak-End Rule
\item Hick's Law

\end{enumerate}
we hypothesise that the animated price drop, if applied with above mentioned constraints, will enhance user engagement, create a sense of anticipation and positively influence purchasing behaviour. 

\section{Literature Review}

\subsection{The Role of Visual Appearance}

Proper use of animation does influence the user in a positive manner. Yet, there are several factors that should be considered when implementing them. One of which is the Aesthetic-Usability effect, which states that proper use of animation to improve visual appearance of the interface will make users more tolerant of minor usability issues(Kate Morgan, 2024). Users tend to love interfaces which are visually appealing. Thus, before the inclusion of any sort of animation in the interface, it becomes extremely important to consider the appearance of the interface. \newline However, along with aesthetics comes usability. How easy is it for users to use a product. A usability test was conducted(Kate Morgan, 2024) in which participants were asked to visit a website(Arcadis, 2024) and present their views. The first thing one of the participants observed was its visual appearance which they stated was appealing. However, usability was one of the major things the participant struggled with thus revising his opinion for which the participant stated that for the first time, it would feel good but the same feeling wouldn't resonate when using it for the second time. \newline The vitality to understand the importance of aesthetics while maintaining usability before implementing any sort of animation is significant. Thus, animation would only end up cluttering the interface if the visual appearance of the interface is not pleasing enough. 

\subsection{User's perception of animation}

The use of several types of user interface animations have been studied in the context of cognitive load and engagement. One such research (Jussi Huhtala, Ari-Heikki Sarjanoja, Jani Mäntyjärvi, Minna Isomursu \& Jonna Häkkilä, 2010) studied users on how they perceive time based on specific types of animated effects on mobile screen which were presented on desktop screens for the purpose of this experiment. The objective was to find out how different timings influence how fast users perceive the animation. The result concluded that early timing(showing the next animation early) is the most effective way of making UI transitions feel faster, enough to create user anticipation and hide delays to improve user satisfaction. The importance of this study is to understand that dynamic animations such as price-drop which will follow after slashing original price to introduce the sale price through "fade-in" animation, should be quick, enough to create user anticipation and making effects feel much smoother. \newline Well made animations will fetch positive results thus driving sales for e-commerce platforms. Such dynamic animations associated with the price display of products creates a deeper impact than the ones with static text. 

\newpage

Users tend to remember the "Peak" and the "end" of an event the most. Hence, a well executed animation contributing to the interface's aesthetics play a huge role in user's subconscious memory. 

What also comes as an obvious fact is that such an animation will reduce the cognitive load on the user because such events act as a visual cue (Wenchao Li, Yun Wang†, Haidong Zhang‡ \& Huamin Qu, 2020) for the users assisting them in identifying sales that matter to them the most. 

\subsection{Challenges and Considerations in using Animation}

While animations do contribute to the look and feel of the interface, it still poses a problem which is associated with specific users such as Autistic users who find animations distracting (Alexandra L. Uitdenbogerd, Maria Spichkova \& Mona Alzahrani, 2022). This study aimed at studying the effect of animation on autistic and non-autistic users. The results though, were significant:

\begin{enumerate}

\item{Even the smallest animations caused stress to Autistic users}
\item{Autistic users tended to enter shorter queries to avoid animations.}

\end{enumerate}
Although, exposure to animation did not affect the user's task performance but affected his/her behaviour. 
Thus, animations must be well thought of, especially when it is to be implemented into e-commerce platforms. It should be minimal, purposeful and non-intrusive while contributing to aesthetics in such a way that it does not increase the cognitive load on the user, particularly for the ones who are more sensitive to visual stimuli.

\section{Theoretical Framework}

\subsection{Theoretical Foundations of Animation in User Interfaces}

The impact of user animation is grounded in several theories of usability, cognitive load and visual aesthetics. The following theories form the foundation of this study:

\subsubsection{Aesthetic-Usability Effect}

The study of users backed by Aesthetic-Usability Effect (Kate Morgan, 2024) suggests that users perceive aesthetically pleasing interfaces more usable, even when minor usability issues exist. The study indicates when visual elements are used in a way to please the human eye, it can enhance user engagement. However, excessive use of unnecessary elements such as animation, buttons, colours or labels can prove harmful for the interface as it would only end up increasing the cognitive load for the user thus affecting the experience tied to the interface. 

\subsubsection{Cognitive-Load Theory}

According to Cognitive Load Theory (Sweller, 1998), users have limited cognitive resources and an overload of animations can impact user performance. Research (Li et al., 2020) shows that dynamic animations serve as visual cues guiding user perception and reducing cognitive load. However, poor use of such resources can only impact the user in a negative manner. 

\subsubsection{Peak-End Rule and User Memory}

The Peak-End Rule states that people tend to remember the peak and the end of an event the most than other aspects of the event. Going by this theory, the primary goal is to execute animations well during key moments such as significant price-drop or discounts on user's wanted products. 

\newpage

\subsection{Conceptual Model for Animation \& Hypotheses}

The key to effectively integrating animations smartly when it comes to niche platforms such as e-commerce is to make sure the execution and the purpose of animations serve well \footnote{The execution of animation is covered in the Proposed Model Section of this paper}. While there exists several ways to use dynamic animations in e-commerce platform, this paper specifically ties its study to price-drop animation as it is one of the major factors that users consider when buying any product. According to the Incentive Theory, external rewards and incentives influence decision-making and behavior (Katie Lundin, 2024). In the context of e-commerce, discounts and sales act as a powerful incentive that can drive user engagement and purchasing behaviour By integrating price-drop animations, platform can leverage this psychological effect to enhance user motivation and shopping decisions. Apart from the factor of delivery time, which tops the "factors enhancing purchase need" chart, sales and discounts are also an intriguing part for the consumer. Thus, integrating animation in price drop doesn't pose anything new, it rather is a refined way of approaching the user mind to enhance the desire to buy a product. \newline As mentioned, animations shouldn't be executed randomly and for anything rather, it should be tied to very specific things such as products the user might be interested in. Likewise, to reduce cognitive load and enhance engagement, products associated with animation shouldn't be tied together as it would only clutter the interface. 

\subsubsection{Hypotheses}

Based on the theories mentioned, and the applied framework the following hypotheses are proposed:

\begin{enumerate}

\item{Aesthetically designed animations will lead to improved engagement and enhanced purchase behaviour}

\item{Well-timed animations (Jussi Huhtala, Ari-Heikki Sarjanoja, Jani Mäntyjärvi, Minna Isomursu \& Jonna Häkkilä, 2010) will create a perception of speed, enhancing user engagement and memory while improving satisfaction}

\item{Excessive use of animations or clustering too many products with the same animated effect on the homepage or search results page will increase cognitive load, negatively impacting user experience.}
\end{enumerate}

\section{Proposed Model}

To effectively integrate animation into e-commerce platforms without overwhelming users, this section proposes a structured, user-centric approach. This model proposes the integration of a dynamic price-drop animation within e-commerce platforms while maintaining an uncluttered interface. The animation is designed to capture user attention towards specific discounted products(in which they might be interested in) without affecting their browsing experience. 

\subsubsection{Homepage Implementation}

Homepage is the very first thing a user views. It becomes extremely important to implement concerned items very strategically. On the homepage, products relevant to the user's interest will be strategically distributed across the page rather than being clustered together. This approach helps in reducing cognitive load, ensuring an interface where distinguishing key elements relevant to the user becomes easier. \newline As the user scrolls and a product enters the viewport, the price-drop animation will be triggered dynamically. The animation sequence will be as follows:

\begin{enumerate}

\item{The original price shifts slightly to the left as a strikethrough effect is applied to it.}
\item{The discounted sale prices appears dynamically, introduced through a fade-in animation.}
\item{The entire transition will occur within 1-1.5 seconds, ensuring it is swift enough to maintain the user's browsing flow while the animation still being noticeable and engaging.}

\end{enumerate}

\newpage

\subsubsection{Search-results Page Implementation}

When it comes to search results page, the user knows what they require. Thus, providing a list of products relevant to the search terms the user used and along with that, providing specific products the user might be interested in based on past purchasing behaviour becomes vital. Thus, as the user clicks the search button, they should be presented with their familiar search results page but with that, a few selected products must be displayed with the animation mentioned in the previous sub-section. 

The rule remains the same, to implement animation in such a way that it does not delay their experience while making them feel engaged. Based on the Peak-End rule, the peak moment is what the user gets to see when they click the search button and the end is how the animation was executed to draw user attention towards that specific product. 
Based on the following theory, users are more likely to remember products associated with animation than those without it.

Then again, for the search results page, products associated with the animation should not be cluttered, rather spread across the page. 

In order to provide a better clearance, following are the images which will consist of 3 frames.

\begin{figure}[h]
    \centering
    \includegraphics[width=0.35\textwidth]{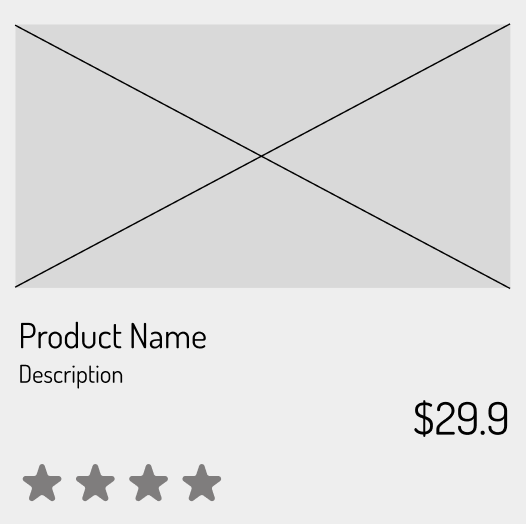}
    \caption{Visual appearance of the product container before animation trigger}
    \label{fig:img1}
\end{figure} 

This figure illustrates how a typical product container appears before the animation begins (t = 0 seconds). At this stage, the product's original price is displayed statically without any visual effects applied.

\begin{figure}[h]
    \centering
    \includegraphics[width=0.35\textwidth]{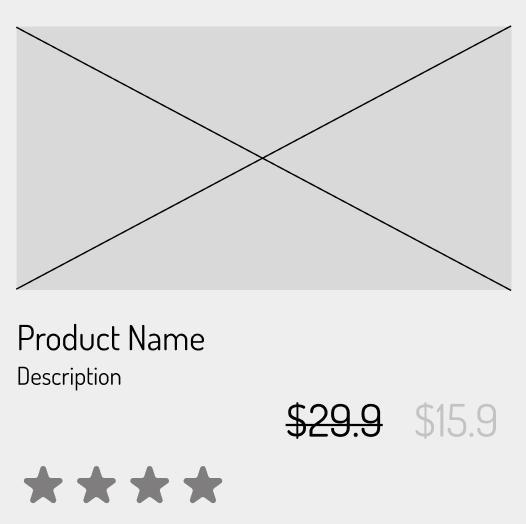}
    \caption{Visual appearance of the product container during animation}
    \label{fig:img2}
\end{figure} 

This figure demonstrates the animation in progress (t = 0.5 seconds). The original price moves slightly to the left while a strikethrough effect is applied. Simultaneously, the discounted sale price starts appearing dynamically through a smooth fade-in effect.

\newpage

\begin{figure}[h]
    \centering
    \includegraphics[width=0.35\textwidth]{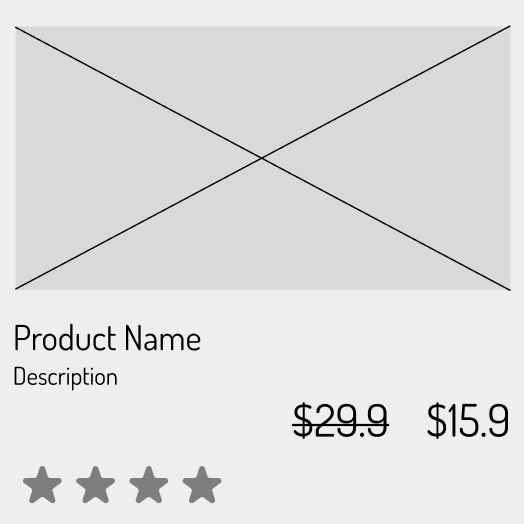}
    \caption{Final product container after animation completes}
    \label{fig:img3}
\end{figure} 
This figure represents the final state of the product container after the animation has finished playing. The interface now appears static, with the strikethrough original price and the newly introduced discounted price. The brief animation enhances user engagement while maintaining a clean and readable design.
\newline But, the implementation of animation before reaching this final static state does make the experience in using the interface much intriguing.

One important factor to consider is that once an animation rolls, it shouldn't keep rolling again and again. From a technical perspective, it'll lead to heavy load on the system and second, from a user perspective, the interest has already been created. Rolling the animation again and again would only frustrate the user.

\section{Discussion and Implication}

\subsection{Discussion}

The integration of dynamic price-drop animations in e-commerce interfaces must be approached with inclusivity in mind. Though animations can enhance user engagement and drive purchasing decisions, their design should consider the diverse cognitive and perceptual needs of users. Research (Alexandra L. Uitdenbogerd, Maria Spichkova \& Mona Alzahrani, 2022) highlights how animations impact autistic and non-autistic users differently, emphasising the importance of designing animations that remain effective without overwhelming users who may be more sensitive to visual stimuli. 

Balancing aesthetic appeal with usability is critical. Overly complex or very fast-paced animations can create cognitive barrier rather than an engaging experience. Therefore, execution of animations should be predictable and non-intrusive. Instead of excessive movement, animations should serve a clear functional purpose, act as guiding cues without overwhelming the user. Next, using familiar, fluid transitions rather than complex or unexpected animations improves accessibility for users having visual processing sensitivities. 

\subsection{Practical Implications for Developers and Designers}

For UI/UX designers and developers, this study states the need to integrate human-centered design principles when implementing animations. There must be a balance between engaging visuals and practical functionality. As demonstrated by a study (Kate Morgan, 2024) usability does concern the user, aesthetics will only create an impression which is independent of how the product is to use. The aim of animation implementation is to ensure users receive a personalised browsing experience. 

Finally, the best way to understand user needs is to conduct user research with individuals who have varying cognitive and sensory needs. This can help refine the usability and interface design of the product. 

\newpage

\section{Limitation and Future Research}

\subsection{Limitations}

While this study provides insights into the role of animations in e-commerce, it also comes with several limitations:

\begin{enumerate}

\item{The study primarily focuses on general user engagement without extensive empirical testing on neurodivergent users such as individuals with autism and sensory sensitivities.}

\item{This study primarily focuses on price-drop animations but the findings may not generalise to other types of UI animations in e-commerce, such as product recommendations, checkout flows, or cart interactions which is a study in itself.}

\item{The study presented is conceptual and theoretical in nature lacking large-scale A/B testing to validate proposed effects of price-drop animations on actual purchase behaviour. This study is only backed by existing theories and experiments.}

\end{enumerate}

\subsection{Future Research Directions}

\begin{itemize}

\item{To address limitations, the future research should focus on conducting large-scale user studies with diverse participants for this study specifically. Using quantitative and qualitative data to understand how animations affect user attention, engagement and decision-making should also be taken into consideration.}

\item{The use of Artificial Intelligence should also be taken into consideration. AI makes a system smart thus, employing the use of AI to explore user preference and browsing history to create a more personalised shopping experience can prove extremely beneficial for an organization. Furthermore, machine learning and AI can be used for UI personalization which can optimise animations, their timings and intensity for individual users.}

\item{Not all devices can incorporate animations and expect to run smoothly without any lag. Thus, there also exists a need for future work to analyze trade-offs between animation complexity and device/website performance.}

\end{itemize}

\section{Conclusion}

This study explored the role of dynamic price-drop animation in e-commerce, particularly in enhancing user engagement and influencing user purchase behaviour. By applying Incentive Theory, this research highlights how well-designed price-drop animations can create a more compelling shopping experience. 

The findings emphasize the importance of strategic product placement associated with animations to avoid cognitive overload while maintaining user interest. Aesthetically pleasing, well-timed and subtle animations can make price reductions more noticeable. However excessive or poorly executed animations can lead to distraction and frustration underscoring the need for balance between usability and aesthetics. 

Additionally, the study acknowledges the necessity of inclusive design in UI animations. Several studies highlight that animations can affect neurodivergent users differently, thus requiring thoughtful implementation to ensure accessibility for a diverse audience. 

Despite this study's insights, the study remains theoretical, lacking large-scale empirical validation. Future research should focus on A/B testing, cognitive load assessments, and AI-driven personalizations to optimize animations even further. 

This study provides a foundation for future empirical research on dynamic animations in e-commerce, highlighting the need for user-centered design in enhancing online shopping experiences.

\end{document}